\documentclass[a4paper,11pt]{article}
\usepackage{latexsym,amssymb}
\begin{document}
\begin{center}{\Large{\bf Model and Statistical Analysis of the Motion of a Tired Random Walker in Continuum}}
\end{center}

\vskip 0.5 cm

\begin{center}{\it Muktish Acharyya}\\
{\it Department of Physics, Presidency University,}\\
{\it 86/1 College Street, Calcutta-700073, INDIA}\\
{E-mail:muktish.physics@presiuniv.ac.in}\end{center}

\vskip 2 cm

\noindent {\bf Abstract:} The model of a tired random walker, whose jump-length
decays exponentially in time, is proposed and the motion of such a tired 
random walker is studied systematically in one, two and
three dimensional continuum. 
In all cases, the diffusive nature of walker, breaks down 
due to tiring which is 
quite obvious. 
In one dimension, the distribution of the displacement of a tired walker
remains Gaussian (as
observed in normal walker) with reduced width. In two and three dimensions, the
probability distribution of displacement becomes nonmonotonic and unimodal. The
most probable displacement and the deviation 
reduces as the tiring factor increases.
The probability of return of a tired walker,
decreases as the tiring factor increases in one and two
dimensions. However, in three dimensions, it is found that the 
probability of return almost insensitive to the tiring factor. 
The probability
distributions of first return time of a tired random walker 
do not show the scale invariance as observed for a normal walker in continuum.
The exponents, of such power law distributions of first return time,
 in all three dimensions are estimated for normal walker.
The exit probability and the probability distribution of first passage time
are found in all three dimensions. A few results are compared with 
available analytical
calculations for normal walker.

\vskip 2cm


\newpage

\noindent {\bf I. Introduction:} 

In statistical physics, 
process of polymerization\cite{rev,grassberger}, diffusion\cite{jkb} 
in restricted geometry etc are some classic
phenomena, which have drawn much attention of the researcher in last
few decades. The underlying mechanism of such physical phenomena 
are tried to explain by
random walk\cite{thesis}. Different types of random walk are studied on the
lattice in different dimensions by the method of computer simulation. 
The absorbing phase transition in a conserved lattice gas with random
neighbour particle hopping is studied\cite{lubeck}. Quenched averages 
for self avoiding walks on random lattices\cite{dd1}, 
asymptotic shape of the region
visited by an Eulerian walker\cite{dd2}, linear and branched avalanches
are studied in self avoiding random walks\cite{manna}, effects of 
quenching is studied in quantum random walk recently\cite{ps}. The drift and
trapping in biased diffusion on disordered lattices
is also studied\cite{stauffer}.

Very recntly, some more interesting results on random walk were reported.
The average number of distinct sites visited by a random walker on the 
random graph\cite{satyada1}, statistics of first passage time of the 
Browian motion, conditioned by maximum value of area\cite{satyada2}
are studied recently.
It may be mentioned here that the first passage time in complex 
scale invariant media was studied\cite{tejedor1}.
The thoery of mean first passage time for jump processes are developed
\cite{tejedor2} and
verified by applying in Levy flights and fractional Brownian motion.
The statistics of the gap and time interval between the highest positions
of a Markovian one dimensional random walker\cite{satyada3}, the universal
statistics of longest lasting records of random walks and Levy flights
are also studied\cite{satyada4}.

The random walks in continuum are studied to model real life problems.
The exact solution of a Brownian inchworm model and self-propulsion
was also studied\cite{sriram}, theory of continuum random walks and 
application in chemotaxis was developed\cite{cont1}. 
Random walks in continuum was
also studied for diffusion and reaction in catalyst\cite{cont2}. 
Very recently, the random walk in continuum is studied with uniformly
distributed random size of flight\cite{aba}. The statistics of Pearson
walk is studied\cite{redner,redner1} in two dimensions for shrinking stepsize
and found a transition of the endpoint distribution by varying the initial
stepsize.

The living random walker in continuum gradually becomes tired as the time
passes, in reality. This would reduce its energy, as a result the 
size of flight gets reduced gradually with time. The first-passage properties
\cite{fpbook}
of a walker is important in various aspects, namely, the fluorescence quenching
in which a fluoresecent molecule stops while reacting with a quencher, firing
neurons when the fluctuating voltage level first reaches a specified value,
in econophysics, the execution of buy/sale orders when a stock price first
reaches a threshold. {\it What will be the first passage properties if the stepsize
of a Pearson walker decreases exponentially in time ?} 
In this paper, 
addressing this particular
problem, 
a model of tired random
walker is proposed in continuum and statistics of its motion is studied systematically
in one, two and three dimensional continuum. The first passage properties, return and
exit probabilities are studied here. The 
numerical results of detailed statistical analysis
of the motion of a tired random walker is also reported here. This paper is organised
as follows: In the next section (section-II) the model of tired random walk is proposed
and the results obtained from numerical simulations are given. The paper ends with a
summary given in section-III.

\newpage

\noindent {\bf II. Model and Results:}

Generally, the motion of a random walker is studied by considering the
time ($t$) independent size ($R$) of flight in each move. 
In this study, the model of a tired
random walker is proposed in such a way that the size of flight of a walker
decreases exponentially as $R(t) = {\rm e^{-\alpha t}}$. 
A simple logic behind it may be stated as follows: if a living cell is moving
continuously, its energy (basically kinetic energy) 
gradually decreases and hence
the velocity, which in turn reduces its size of flight 
(i.e., jump-length per unit
time).
Here, $\alpha$ is tiring
factor. The statistical behaviour of such a tired random walker is studied
in one, two and three dimensional continuum. It may be noted here that such
kind of behaviour 
of a tired random walker cannot be studied on the lattice.

In one dimension, the size of flight in each time step is 
$x_l(t)={\rm e^{-\alpha t}}$. A walker starts its journey from the origin having
the equal probability of choosing the left and right direction. 
The updating rule, in one dimemsional tired walk, 
 may be expressed as: $x(t+1) = x(t) \pm x_l(t)$

In two
dimensions (planar continuum), the tired walker starts its journey from 
the origin and it has a uniform probability of choosing any random direction
($\theta$) distributed between 0 and $2\pi$. Its motion can be
represented mathematically as:

\begin{equation}
x(t+1) = x(t) + R(t)cos(\theta); 
y(t+1) = y(t) + R(t)sin(\theta)
\end{equation}

\noindent The displacement at time $t$ is $r(t) = \sqrt{x^2(t) +y^2(t)}$.
In planar continuum, the area of the region visited 
by a tired walker is obviously
shorter than that visted by a normal walker, in a specified 
course of time. A typical such comparison
is shown in Fig-1 with $\alpha=0.001$. As a result, the mean square
displacement does not show diffusive behaviour as shown by a normal walker.
In long time, it gets saturated (motion stops practically). 

A typical such comparison is shown in
Fig-2 for $\alpha=0.001$ and $\alpha=0.0005$. 
The similar behaviours are also observed in one
and three dimensions (not shown). The tired walk is not diffusive ($<r^2> \sim t$)
as observed in normal ($\alpha=0$) walk.
It is also observed that the motion stops earlier if the tiring factor $\alpha$
increases.

Now, let it be discussed systematically in one, two and three dimensions.
In one dimensions, the probability distribution of the displacements 
of a walker are
studied for $\alpha = 0.0$ (normal), 0.001(moderately tired) and 
0.01(heavily tired). As usual, the distribution is normal (Gaussian)
with zero mean in 
all the cases. However, as the tiring factor 
increases the distribution becomes
sharper and sharper. These are depicted in Fig-3. Here, it may be mentioned that
the values of $\alpha$ and the maximum time allowed ($N_t$) are such that the
walker gets frozen (due to exponential decrease of step-size after such long
time). The distribution shown in Fig-3, is practically the density distribution
of frozen walker. It would be interesting to study the density 
distribution of these frozen
walker as a function of $\alpha$ through the scaling.

What will be the probability of return($P_R$)
 in one dimension ? First of all, in
continuum one should be careful in defining the probability of return. In the
lattice the probability of return is defined as the walker returns to its
initial starting point. However, in continuum, it is quite unlikely that a
tired walker returns to its initial starting point. Here, one may think that
whether the tired walker returns within a linear zone ($[-r_z,r_z]$)
centered
around the origin. Now put a large number ($N_s$) of walker at origin and
allow them to walk (with different random sequence) 
upto a certain time ($N_t$) 
and then check how many walkers return within the preassigned
returning zone (of size $r_z$). The calculated fraction 
is the probability of return (within time of observation $N_t$) in
this particular model. In the lattice model this probability is 1, which can
also be derived from exact calculations\cite{poyla}. In this model of tired walker, 
considering $r_z=0.5$, 
$P_R = 0.992$ for $\alpha=0.0$. This numerical estimate of returm probability agrees well with exact
calculation of return probability ($P_R = 1$)\cite{poyla} in one dimensional
normal random walk. It may be noted here that for $\alpha=0$, the walker 
returns at origin (the starting point also) and the probability of return
can be compared to that obtained in random walk on one dimensional lattice. 
As the tiring factor increases, $P_R$
decreases. For moderately tired ($\alpha=0.001$) walker, $P_R=0.955$ and for
heavily tired walker ($\alpha=0.01$), $P_R=0.874$. In Fig-4, the $P_R$ is
plotted against $N_t$ for various values of $\alpha$. Now, this probability
of return ($P_R$) must depend on the size ($r_z$) of returning zone. To 
study the dependences of $P_R$ on $r_z$, $P_R$ is studied as
a function of $r_z$ for
different values of $\alpha$ and shown in Fig-5. It shows that the $P_R$ 
grows as $r_z$ increases in the case of tired walker ($\alpha=0.001, 0.01$),   
but $P_R$ does not depend on $r_z$ for normal walker. 
It is important to note here
that even for heavily tired ($\alpha=0.01$) random walker, the size of the
flight, after $t=10$ is larger than 0.90. So, 
the range of values of $r_z$, chosen here, does not have any
chance that the walker remains in the returning zone immediately after 
starting its journey.  So, the choice $r_z$ =0.5 is quite safe to
study the probability of return in this context.

How long a tired walker takes to return first 
time within returing zone ?
How does the distribution of this first returning time ($t_r$) look like ?
The probability distribution of first returning time ($t_r$) of a tired
random walker is shown in Fig-6. A normal walker ($\alpha=0.0$) shows a 
scale invariant ($P(t_r) \sim t_r^{-\beta}$) 
distribution of first returning
time ($t_r$). The exponent estimated is $\beta \simeq 1.49$. This result agrees well with 
analytical result \cite{fpbook}, where it is found $P(t_r) \sim t_r^{-3/2}$. 
However, this scale invariant nature
of the distribution of first returning time, 
breaks down in the cases of tired
walking (for $\alpha=0.001, 0.01$) (see Fig-6). More detail investigation is required to
propose any functional behaviour of $P(t_r)$ for $\alpha \neq 0$.

In one dimension, how long ($t_e$) a tired walker takes to exit (first time) from a zone ($[-r_e,r_e]$) ?  
The probability distribution ($P(t_e)$) of first passage time ($t_e$) (for a fixed
value of $r_e=25.0$), is studied for different values of $\alpha$
and shown in Fig-7(a). As the $\alpha$ increases, 
the most probable first passage
time decreases. It should be noted here that the probability of first 
passage is defined (in this study) as the probability 
to escape in a given time, 
from a bounded ($[-r_e,r_e]$) linear (in one dimension) region.
If it would be defined as the probability to escape through a given point
(say $x=r_e$) the power law ($P(t_e) \sim t_e^{-1.49}$) distribution 
in long time limit ($t_e \to \infty$) is found
which supports the analytical prediction 
($P(t_e) \sim t_e^{-3/2}$)\cite{fpbook}. This is shown in Fig-7(b).

What is the probability of exit $(P_e)$ of a tired walker one dimensional
continuum ?
The exit probability (for a fixed time of observation $N_t$)
from a zone of absolute distance $r_e$ (measured from the origin)
is also studied, in one dimension, as a function 
of $r_e$ and shown in Fig-8. Here, the exit
probability, of a tired walker, was found to decreases as the 
absolute distance of zone ($r_e$) increases. However, it remains fixed 
(nearly 1) for
a normal walker\cite{poyla}. It may also be noted that, the rate of fall of exit
probability increases as the tiring factor ($\alpha$) increases.

In two dimensions, the motion of a tired random walker is studied by using
the rule given in equation -1. Here, the mean square displacement $<r^2>$ is 
proportinal to the time $t$ for $\alpha=0.0$, reveals the 
conventional diffusive ($<r^2> \sim t$) behaviour.
However, a moderately tired ($\alpha=0.001 \& 0.0005$) walker does not show long time
diffusive behaviour. This is quite obvious and already shown in Fig-2. 

The distribution of absolute displacement is nonmonotonic unimodal function.
It is shown in Fig-9. It is observed that the maximum probability of finding
the walker at a distance ($r_m$) 
from the origin and the average distance ($\bar r$)
both decreases as the tiring factor ($\alpha$) increases. In this case,
$r_m=64.0$, $\bar r = 89.13$ for $\alpha=0.0$, $r_m=15.0$, 
$\bar r = 20.04$ for $\alpha=0.001$ and  $r_m=5.0$, $\bar r=6.42$ for $\alpha=0.01$.  
Here also this distribution is practically the density distribution of frozen 
walker.

What will be the probability of return ($P_R$) of a tired walker 
in planar continuum ? 
The probability of return within a circle of return having radius $r_z=0.5$
is studied as a function of maximum time of observation $N_t$ and 
shown in Fig-10. In planar continuum, a tired walker has a probability of
return in a circle of radius $r_z=0.5$ as follows: for $\alpha=0.0$, 
$P_R=0.737$, for $\alpha=0.001$, $P_R=0.620$ and for $\alpha=0.01$, 
$P_R=0.524$.

For a fixed value of $N_t$, the probability of return of a tired walker in
planar contunuum, grows as the radius of returning zone increases. This
is shown in Fig-11.

Here, like in one dimensional tired walker, the probability distribution
of first returning time $(t_r)$ shows a scale invariance
($P(t_r) \sim t_r^{-\beta}$) for $\alpha=0.0$
with $\beta \simeq 1.10$. 
However, the analytic result\cite{fpbook} suggests $P(t_r) \sim 1/(t_r (ln(t_r)^2)$.
The possible reason of disagreement may be stated as follows: 
in the analytic calculation
of $P(t_r)$, it was defined as the 
probability of return exactly at the origin from where the
walker has started its journey. However, in the numerical 
simulation, $P(t_r)$ is defined as the probability
of return (first time) within a circular zone of radius $r_z$.
As the tiring factor ($\alpha \neq=0$)
increases, the scale invariance 
nature of the distribution on first returning time
breaks down. This is demonstrated in Fig-12.

In two dimensions, the distribution of first passage time (for a fixed
distance $r_e=25.0$), is studied for different values of $\alpha$
and shown in Fig-13. As the $\alpha$ increases, the most probable first passage
time and mean first passage time decreases.

The exit probability (for a fixed time of observation $N_t$)
from a circular zone of radius $r_e$ (measured from the origin)
is also studied, in two dimensions, as a function 
of $r_e$ and shown in Fig-14. Here, the exit
probability was found to decreases as the radius of circular
zone ($r_e$) increases. Here also, the rate of fall of exit probability
increases as the tiring factor ($\alpha$) increases. However, the
exit probability of a normal walker ($\alpha=0$) remains unchanged
(nearly 1) as $r_e$ increases.

The tired walk in three dimensional continuum can also be generalized.
The updating of coordinates obey the following rule:

\begin{flushleft}
$x(t+1)=x(t)+R(t)sin(\theta)cos(\phi)$; \
$y(t+1)=y(t)+R(t)sin(\theta)sin(\phi)$\end{flushleft}
\begin{equation}
z(t+1)=z(t)+R(t)cos(\theta)
\end{equation}

\noindent where $R(t)=e^{-\alpha t}$, $\theta$ is uniformly distributed
random angle between 0 and $\pi$ and $\phi$ is uniformly distributed random
angle between 0 and $2\pi$. The displacement at time $t$ is 
$r(t)=\sqrt{x^2(t) +y^2(t)+z^2(t)}$.

In 3D continuum, the motion of a tired random walker is studied by using
the rule given in equation -2. Here, the mean square displacement $<r^2>$ is 
proportinal to time $t$ for 
$\alpha=0.0$, reveals the diffusive behaviour (not shown).
However, a tired ($\alpha \neq 0$) walker does not show long time
diffusive behaviour (not shown). 

The probability distribution of absolute displacement 
(or the density distribution of frozen walker in reality)
in 3D continuum is 
observed to be a nonmonotonic unimodal function.
It is shown in Fig-15. It is observed that the maximum probability of finding
the walker at a distance ($r_m$) 
from the origin and the mean displacement ($\bar r$)
both decreases as the tiring factor ($\alpha$) increases. In this case,
$r_m=72.0$, $\bar r = 91.59$ for $\alpha=0.0$, $r_m=16.0$, $\bar r = 20.65$ 
for $\alpha=0.001$ and  $r_m=3.0$, $\bar r=6.96$ for $\alpha=0.01$.  

What will be the probability of return in 3D continuum ? 
The probability of return within a sphere of return having radius $r_z=0.5$
is studied as a function of maximum time of observation $N_t$ and 
shown in Fig-16. In 3D continuum, unlike the cases in 1D and 2D continuum,
a tired walker has a probability of
return in a sphere of radius $r_z=0.5$ is almost insensitive ($P_R=0.226$)
of the tiring
factor $\alpha$. 

For a fixed value of $N_t$, the probability of return of a tired walker in
3D contunuum, grows as the radius of returning zone increases keeping
the independence on tiring factor $\alpha$. This
is shown in Fig-17.

In 3D continuum, the probability distribution 
of first returning time $(t_r)$ shows a scale invariance
($P(t_r) \sim t_r^{-\beta}$) for $\alpha=0.0$. 
The exponent estmated $\beta \simeq 1.48$. Accidentally, this is close
the analytical prediction ($P(t_r) \sim t_r^{-3/2}$)\cite{fpbook}.
As the tiring factor ($\alpha \neq 0$)
increases, the scale invariance of the distribution on first returning time,
breaks down. This is demonstrated in Fig-18.

In three dimensions, the distribution of first passage time (for a fixed
distance $r_e=25.0$), is studied for different values of $\alpha$
and shown in Fig-19. As the $\alpha$ increases, the most probable first passage
time and the mean first passage time  decreases.

The exit probability (for a fixed time of observation $N_t$)
from a spherical zone of radius $r_e$ (measured from the origin)
is also studied, in two dimensions, as a function 
of $r_e$ and shown in Fig-20. Here, the exit
probability, of a tired walker, 
was found to decreases as the radius of circular
zone ($r_e$) increases. However, like the earlier cases, it reamins
same (nearly 1) for all $r_e$. Here also the rate, of fall of the
exit probability of a tired walker, increases as the tiring factor
$\alpha$ increases.

\newpage


\vskip 1cm

\noindent {\bf III. Summary:}

In this article, a model of tired random walker in continuum is proposed.
Generally, a random walker moves with constant size of flight. However, as
the time passes, if the walker gets tired, one should think of a time
dependent size of flight. Here, this size of flight decays exponentially
with time. The motion of such a tired walker is studied in one, two and
three dimensional continuum. In this statistical investigation, the 
distribution of the absolute displacement, mean displacement, probability
of return (within a specified zone), distribution of time of first return
are studied systematically. In one and two dimensional continuum, the 
probability of return decreases as the tiring factor increases. However,
in three dimensional continuum, this probability of return seems to be
independent of the tiring factor. The distribution of first returning time
in all dimensions (for normal walker with tiring factor $\alpha=0$), shows
power law behaviours. This scale invariance of the distribution of first
returning time breaks down for $\alpha \neq 0$ in all dimensions. In the 
study of first returning probability, a very important point should be
mentioned. For $\alpha=0$, the probability of return could be compared
with that calculated analytically\cite{fpbook} 
in one dimension only, where the walker
can return to the initial point. In higher dimensions, it returns within
a circular (spherical) zone in two (three) dimensions.

The exit probability and the distribution of first passage time are studied.
In all dimensions, the exit probability was found to decrease as the size
of the zone (from where the tired walker exits out) increases. The rate
of decrease of the exit probability was found to increase as the tiring factor
$\alpha$ increases. Here also the probability of 
first passage (for $\alpha=0$) can only be
compared with analytical calculations\cite{fpbook} in one dimension, if it 
is defined as the probability of escape through a particular point.

The first passage time is defined (in this simulational study)
as the time required by a walker to exit
from a specified zone. This time has a distribution and this distribution is
studied for various values of $\alpha$. It was observed, in all dimensions,
the most probable first passage time decreases as $\alpha$ increases. A rigorous
analysis and possible scaling behaviour (if any) may be investigated.  

Some more interesting studies can be done in this field. In this paper, only
the numerical results are reported. A rigorous mathematical formulations
of first passage properties for {\it tired walk}
has to be developed following the same already developed\cite{fpbook} for
normal walk ($\alpha=0$). 

The possibilities of scaling of distribution of return time, distribution of 
first passage time, distribution of distances and exit probabilities with 
respect to the tiring factor ($\alpha$) has also to be explored.

\vskip 1cm

\noindent {\bf Acknowledgements:} Author would like to express sincere
gratitudes to D. Dhar, S. S. Manna and P. Sen for important discussions.
The library facilities of Calcutta University is gratefully acknowledged.

\newpage
\setlength{\unitlength}{0.240900pt}
\ifx\plotpoint\undefined\newsavebox{\plotpoint}\fi
\sbox{\plotpoint}{\rule[-0.200pt]{0.400pt}{0.400pt}}%


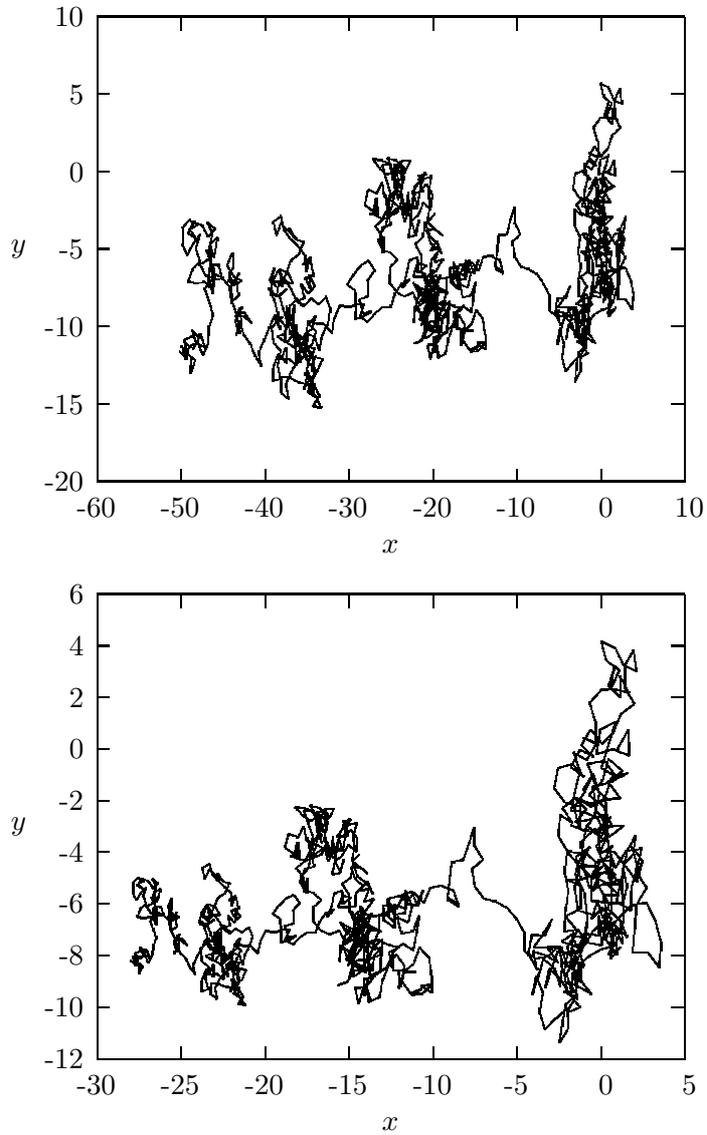
\noindent {\bf Fig-1.} A sample random walk in 2D continuum. The top one shows 
a walk for $\alpha=0.0$ and the bottom one shows a tired 
walk for $\alpha=0.001$. 
Here, in both figures same sequence of random numbers are used.
Note the scales of the axes of two figures. $N_t=10^3$ in both cases.

\newpage
\setlength{\unitlength}{0.240900pt}
\ifx\plotpoint\undefined\newsavebox{\plotpoint}\fi
\sbox{\plotpoint}{\rule[-0.200pt]{0.400pt}{0.400pt}}%
%

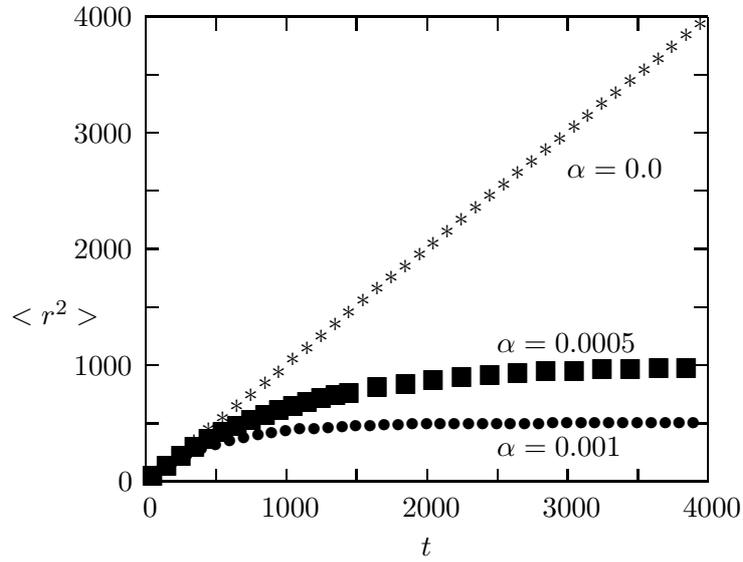
\noindent {\bf Fig-2.} The mean square displacement ($<r^2>$) versus 
time ($t$) in two dimensions for
various values of $\alpha$ (marked by different symbols). The
tired walkers ($\alpha \neq 0$) do not show the 
normal ($<r^2> \sim t$) diffusive behaviour.

\newpage

\setlength{\unitlength}{0.240900pt}
\ifx\plotpoint\undefined\newsavebox{\plotpoint}\fi
\sbox{\plotpoint}{\rule[-0.200pt]{0.400pt}{0.400pt}}%
%

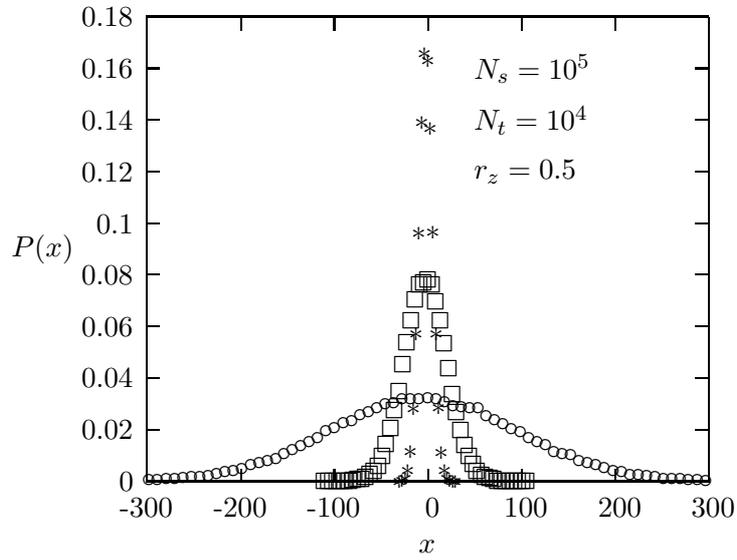
\noindent {\bf Fig-3.} The distribution ($P(x)$) of the displacements ($x$) of
a tired random walker in one dimension. $(\circ)$ represents $\alpha=0.0$, 
$(\Box)$ represents $\alpha=0.001$,
and $(\ast)$ represents $\alpha=0.01$. Note that tiring reduces the width
of the distribution.

\newpage
\setlength{\unitlength}{0.240900pt}
\ifx\plotpoint\undefined\newsavebox{\plotpoint}\fi
\sbox{\plotpoint}{\rule[-0.200pt]{0.400pt}{0.400pt}}%
%

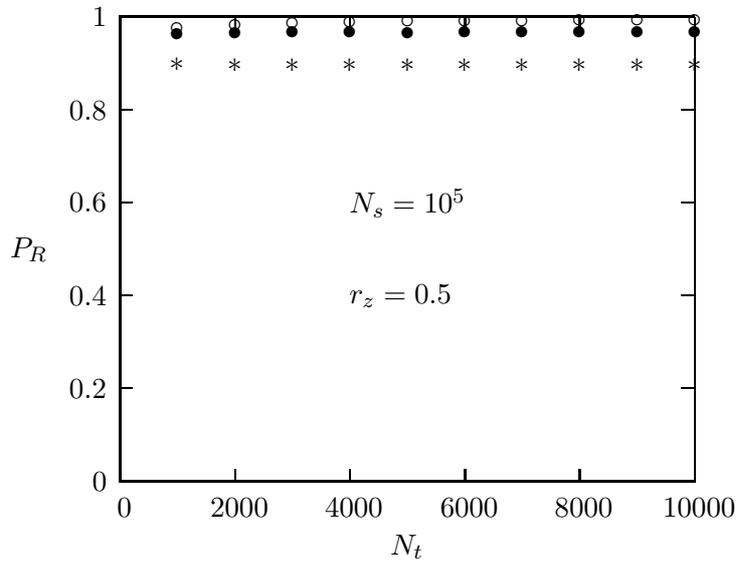
\noindent {\bf Fig-4.} Probability of return 
($P_R$)
plotted against the maximum
time $N_t$ in one dimension.
Different symbols correspond to different values of $\alpha$.
$\alpha=0.0 (\circ)$, $\alpha = 0.001(\bullet)$ and $\alpha=0.01(\ast)$. 
Here in all cases
$N_s=10^5$. The absolute distance ($r_z$) of returning zone 
($[-r_z,r_z]$) is $r_z=0.5$ here.

\newpage
\setlength{\unitlength}{0.240900pt}
\ifx\plotpoint\undefined\newsavebox{\plotpoint}\fi
\sbox{\plotpoint}{\rule[-0.200pt]{0.400pt}{0.400pt}}%


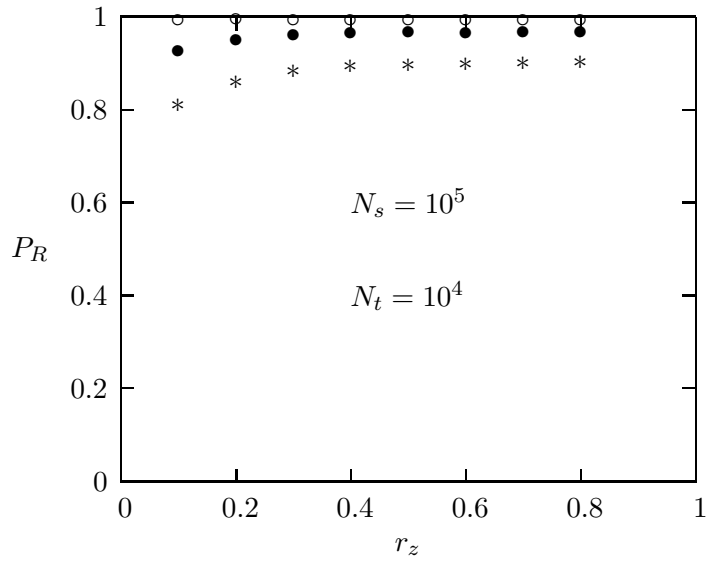
\noindent {\bf Fig-5.} 
Probability of return 
($P_R$)
plotted against the absolute distance $r_z$
of returning zone ($[-r_z,r_z]$)
in one dimension.
Different symbols correspond to different values of $\alpha$.
$\alpha=0.0 (\circ)$, $\alpha = 0.001(\bullet)$ and $\alpha=0.01(\ast)$. 

\newpage
\setlength{\unitlength}{0.240900pt}
\ifx\plotpoint\undefined\newsavebox{\plotpoint}\fi
\sbox{\plotpoint}{\rule[-0.200pt]{0.400pt}{0.400pt}}%
%

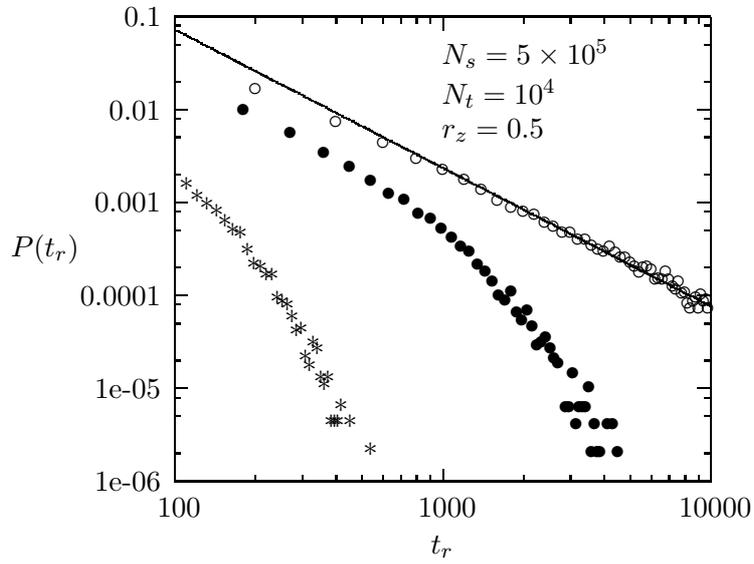
\noindent {\bf Fig-6.} Probability distribution ($P(t_r)$) of first returning 
time ($t_r$) in one dimension.
Different symbols correspond to different values of $\alpha$.
$\alpha=0.0 (\circ)$, $\alpha = 0.001(\bullet)$ and $\alpha=0.01(\ast)$. 
The solid line
is $y=70.0x^{-1.49}$.

\newpage
\setlength{\unitlength}{0.240900pt}
\ifx\plotpoint\undefined\newsavebox{\plotpoint}\fi
\sbox{\plotpoint}{\rule[-0.200pt]{0.400pt}{0.400pt}}%
%

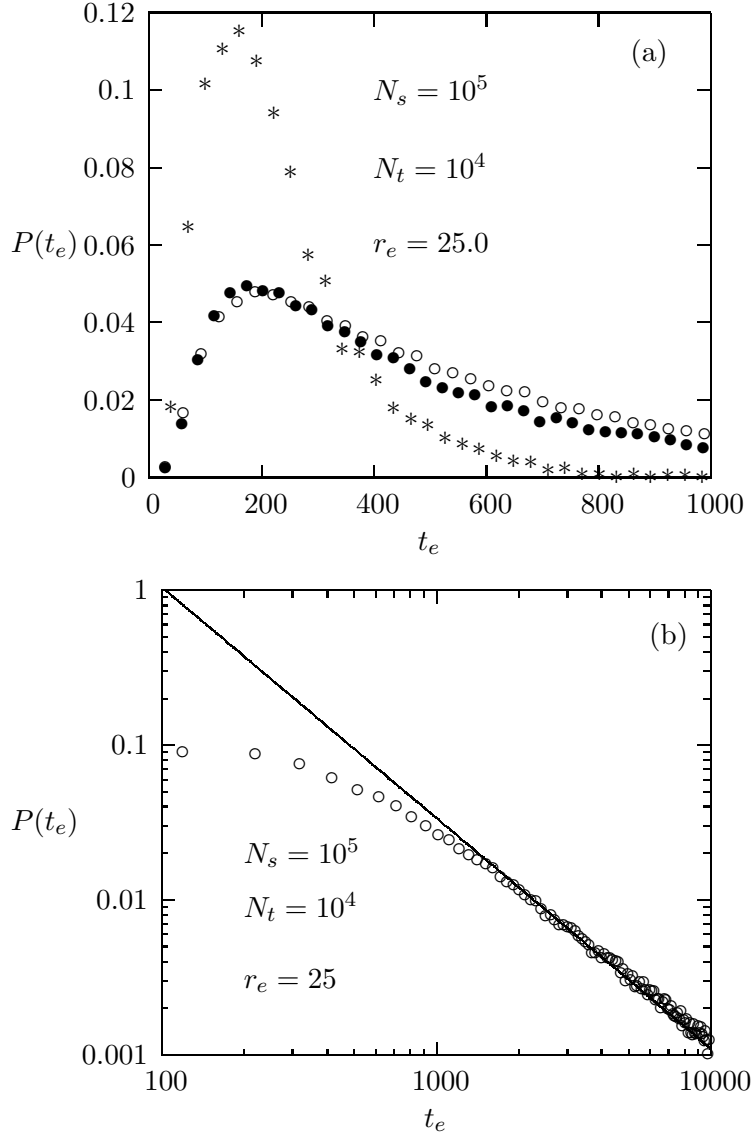
\noindent {\bf Fig-7.} Probability distribution ($P(t_e)$)of 
first passage time ($t_e$) in one dimension.
(a) Different symbols correspond to different values of $\alpha$.
$\alpha=0.0 (\circ)$, $\alpha = 0.001(\bullet)$ and $\alpha=0.004(\ast)$. 
Here, $N_s=10^5$, $N_t=10^4$ and $r_e=25.0$. In (a) the first passage
is defined as the probability 
of exit first from a bounded ($[-r_e,r_e]$) linear region around the origin. 
(b) the first
passage is defined to cross first a point (here $r_e=+25.0$). The solid line
is $y=1000x^{-1.49}$ supporting analytical prediction
($P(t_e) \sim t_e^{-1.5}$)\cite{fpbook}.

\newpage
\setlength{\unitlength}{0.240900pt}
\ifx\plotpoint\undefined\newsavebox{\plotpoint}\fi
\sbox{\plotpoint}{\rule[-0.200pt]{0.400pt}{0.400pt}}%


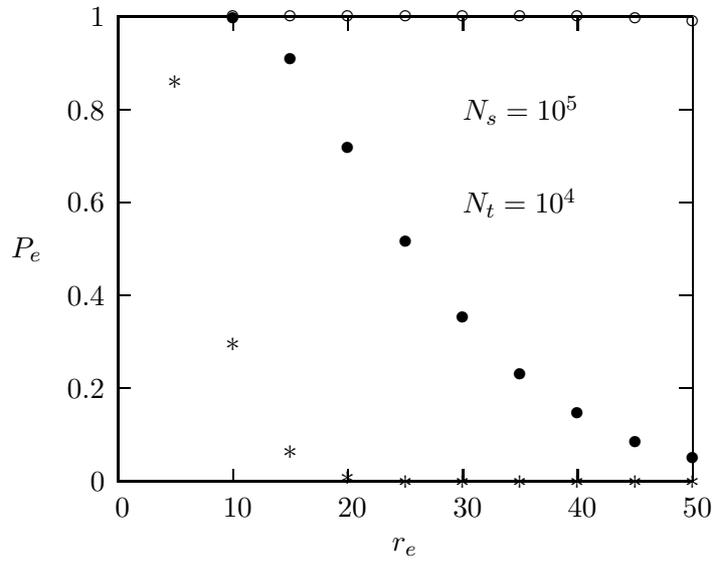
\noindent {\bf Fig-8.} Exit probability 
$P_e$ plotted against $r_e$ in one dimension.
Different symbols correspond to different values of $\alpha$.
$\alpha=0.0 (\circ)$, $\alpha = 0.001(\bullet)$ and $\alpha=0.01(\ast)$. 
Here, $N_s=10^5$ and $N_t=10^4$.

\newpage
\setlength{\unitlength}{0.240900pt}
\ifx\plotpoint\undefined\newsavebox{\plotpoint}\fi
\sbox{\plotpoint}{\rule[-0.200pt]{0.400pt}{0.400pt}}%
%

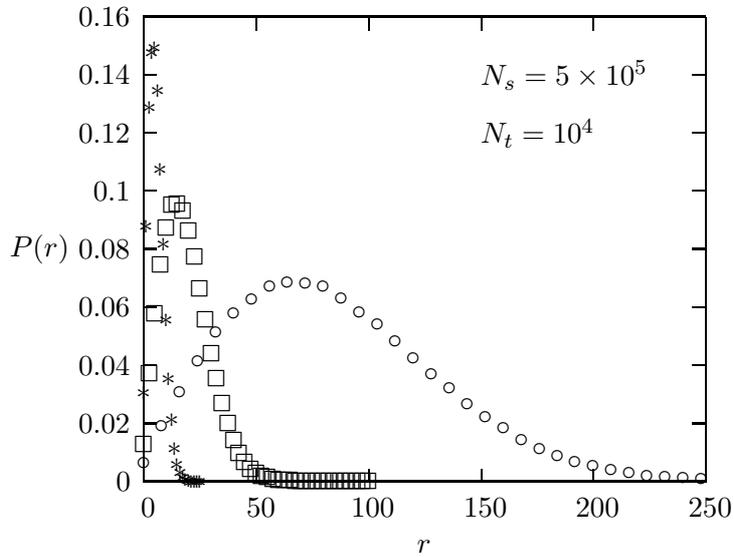
\noindent {\bf Fig-9.} The distribution of  
displacements (${r}$) of
a tired random walker in two dimensions. $(\circ)$ represents $\alpha=0.0$, 
$r_m=64.0$, ${\bar r}$=89.13, $(\Box)$ represents $\alpha=0.001$,
$r_m=15.0$, ${\bar r}$=20.04 and $(\ast)$ represents $\alpha=0.01$,
$r_m=5.0$, ${\bar r}$=6.42. 

\newpage
\setlength{\unitlength}{0.240900pt}
\ifx\plotpoint\undefined\newsavebox{\plotpoint}\fi
\sbox{\plotpoint}{\rule[-0.200pt]{0.400pt}{0.400pt}}%
%

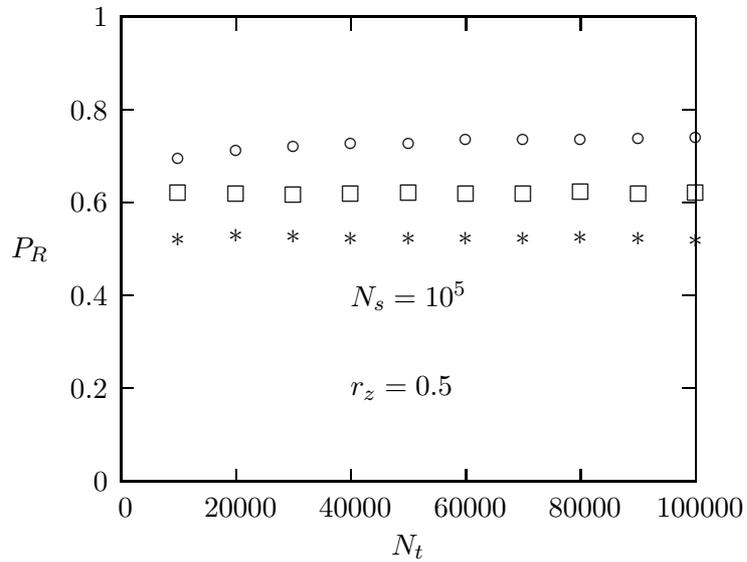
\noindent {\bf Fig-10.} Probability of return 
($P_R$) plotted against $N_t$ in
two dimensions. 
Different symbols correspond to different values of $\alpha$.
$\alpha=0.0 (\circ)$, $\alpha = 0.001(\Box)$ and $\alpha=0.01(\ast)$. 
Here in all cases
$N_s=10^5$. The radius of circular returning zone is $r_z=0.5$.

\newpage
\setlength{\unitlength}{0.240900pt}
\ifx\plotpoint\undefined\newsavebox{\plotpoint}\fi
\sbox{\plotpoint}{\rule[-0.200pt]{0.400pt}{0.400pt}}%
%

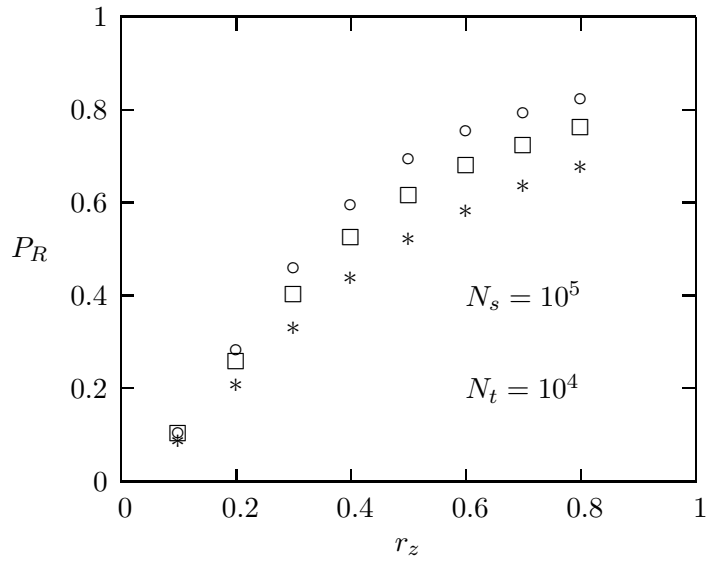
\noindent {\bf Fig-11.}
Probability of return 
($P_R$)
plotted against the radius
of returning zone ($r_z$)
in two dimensions.
Different symbols correspond to 
different values of $\alpha$.
$\alpha=0.0 (\circ)$, $\alpha = 0.001(\Box)$ and $\alpha=0.01(\ast)$. 

\newpage
\setlength{\unitlength}{0.240900pt}
\ifx\plotpoint\undefined\newsavebox{\plotpoint}\fi
\sbox{\plotpoint}{\rule[-0.200pt]{0.400pt}{0.400pt}}%
%

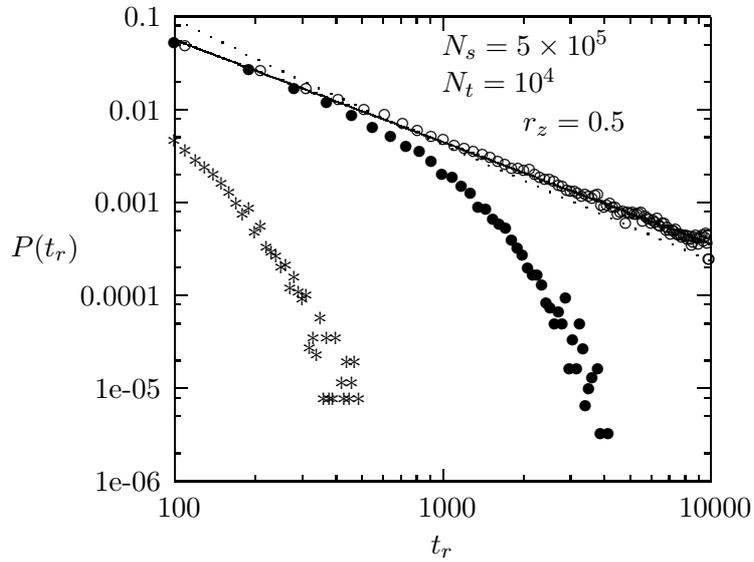
\noindent {\bf Fig-12.} Probability distribution ($P(t_r)$)of first 
returning time ($t_r$)
of two dimensions. 
Different symbols correspond to different values of $\alpha$.
$\alpha=0.0 (\circ)$, $\alpha = 0.001(\Box)$ and $\alpha=0.01(\ast)$. 
Solid line represents $y=9x^{-1.10}$. 
The dotted line is $y=200/(x (ln(x))^2)$\cite{fpbook}. 

\newpage
\setlength{\unitlength}{0.240900pt}
\ifx\plotpoint\undefined\newsavebox{\plotpoint}\fi
\sbox{\plotpoint}{\rule[-0.200pt]{0.400pt}{0.400pt}}%
%

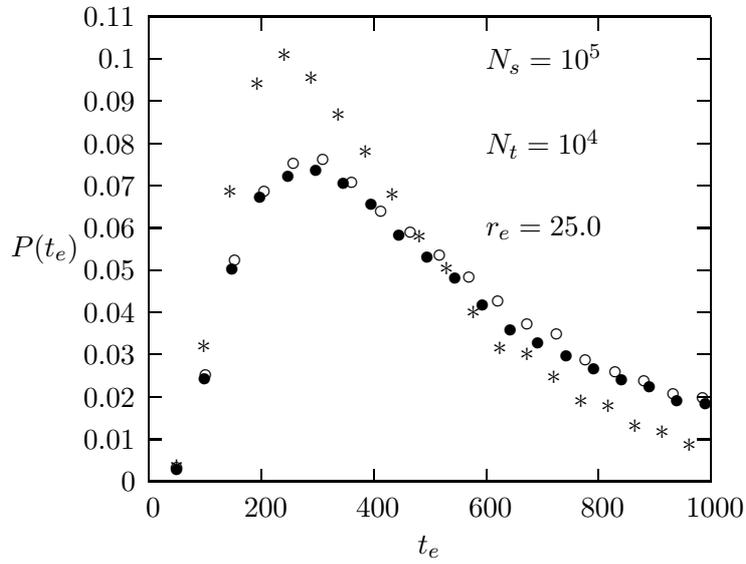
\noindent {\bf Fig-13.} Probability distribution ($P(t_e)$) of 
first passage time ($t_e$) in two dimensions.
Different symbols correspond to different values of $\alpha$.
$\alpha=0.0 (\circ)$, $\alpha = 0.001(\bullet)$ and $\alpha=0.002(\ast)$. 
Here, $N_s=10^5$, $N_t=10^4$ and $r_e=25.0$.

\newpage
\setlength{\unitlength}{0.240900pt}
\ifx\plotpoint\undefined\newsavebox{\plotpoint}\fi
\sbox{\plotpoint}{\rule[-0.200pt]{0.400pt}{0.400pt}}%
%

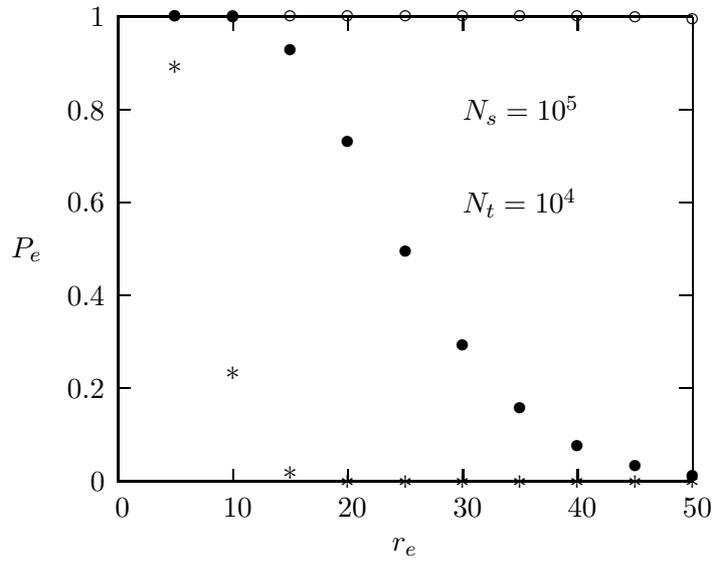
\noindent {\bf Fig-14.} Exit probability 
$P_e$ plotted against $r_e$ in two dimensions.
Different symbols correspond to different values of $\alpha$.
$\alpha=0.0 (\circ)$, $\alpha = 0.001(\bullet)$ and $\alpha=0.01(\ast)$. 
Here, $N_s=10^5$ and $N_t=10^4$.

\newpage
\setlength{\unitlength}{0.240900pt}
\ifx\plotpoint\undefined\newsavebox{\plotpoint}\fi
\sbox{\plotpoint}{\rule[-0.200pt]{0.400pt}{0.400pt}}%
%

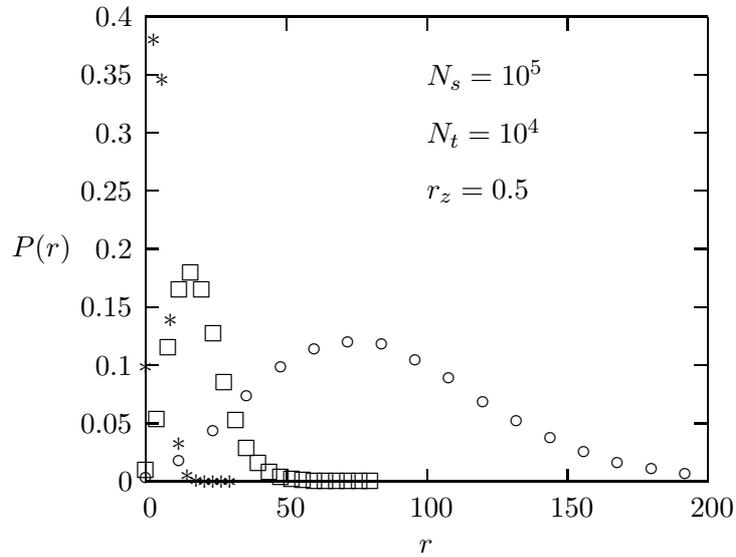
\noindent {\bf Fig-15.} 
The distribution of the displacements of
a tired random walker in three dimensions. $(\circ)$ represents $\alpha=0.0$, 
$r_m=72.0$, ${\bar r}$=91.59, $(\Box)$ represents $\alpha=0.001$,
$r_m=16.0$, ${\bar r}$=20.65 and $(\ast)$ represents $\alpha=0.01$,
$r_m=3.0$, ${\bar r}$=6.96. 

\newpage

\newpage
\setlength{\unitlength}{0.240900pt}
\ifx\plotpoint\undefined\newsavebox{\plotpoint}\fi
\sbox{\plotpoint}{\rule[-0.200pt]{0.400pt}{0.400pt}}%
%

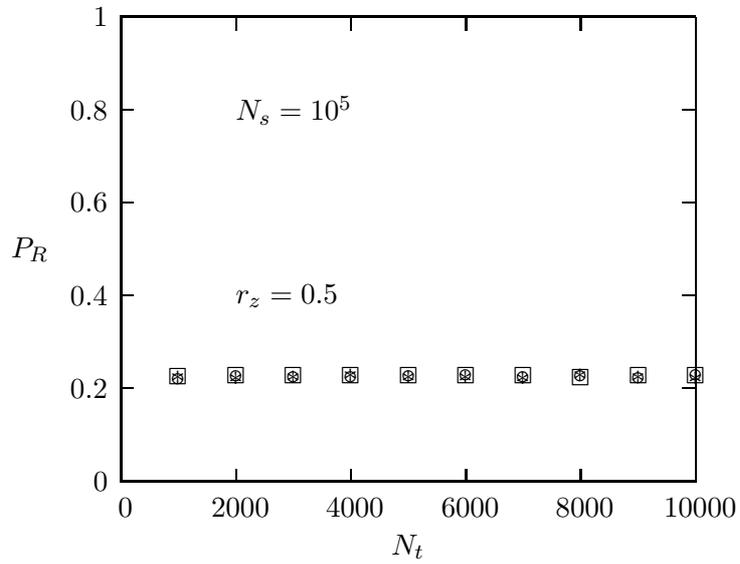
\noindent {\bf Fig-16.} Probability of return ($P_R$) 
versus $N_t$ in three dimensions.
Different symbols correspond to different values of $\alpha$.
$\alpha=0.0 (\circ)$, $\alpha = 0.001(\Box)$ and $\alpha=0.01(\ast)$. 
Here in all cases
$N_s=10^5$. The radius of spherical returning zone is $r_z=0.5$.

\newpage
\setlength{\unitlength}{0.240900pt}
\ifx\plotpoint\undefined\newsavebox{\plotpoint}\fi
\sbox{\plotpoint}{\rule[-0.200pt]{0.400pt}{0.400pt}}%
%

\noindent {\bf Fig-17.} Probability of return ($P_R$) versus $r_z$ 
in three dimensions.
Different symbols correspond to 
different values of $\alpha$.
$\alpha=0.0 (\circ)$, $\alpha = 0.001(\Box)$ and $\alpha=0.01(\ast)$. 

\newpage
\setlength{\unitlength}{0.240900pt}
\ifx\plotpoint\undefined\newsavebox{\plotpoint}\fi
\sbox{\plotpoint}{\rule[-0.200pt]{0.400pt}{0.400pt}}%
%

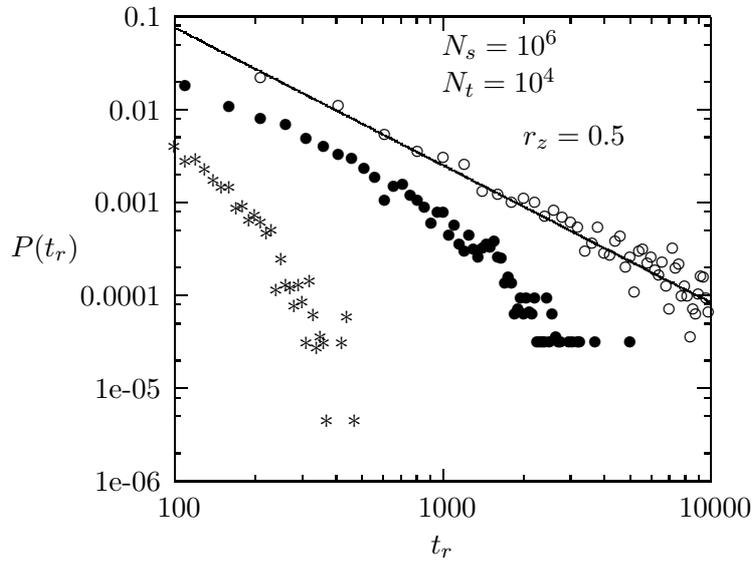
\noindent {\bf Fig-18.} Probability distribution ($P(t_r)$) of 
first returning time ($t_r$) 
in three dimensions.
Different symbols correspond to different values of $\alpha$.
$\alpha=0.0 (\circ)$, $\alpha = 0.001(\Box)$ and $\alpha=0.01(\ast)$. 
Solid line represents $y=70x^{-1.48}$.

\newpage
\setlength{\unitlength}{0.240900pt}
\ifx\plotpoint\undefined\newsavebox{\plotpoint}\fi
\sbox{\plotpoint}{\rule[-0.200pt]{0.400pt}{0.400pt}}%
%

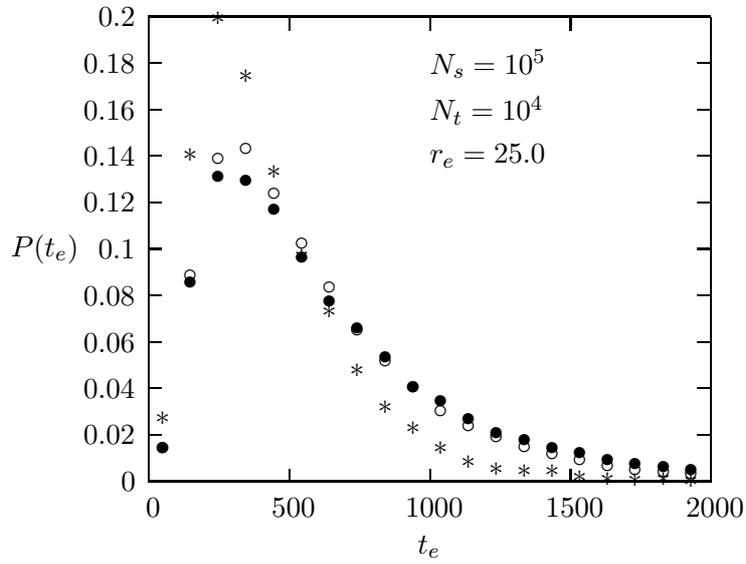
\noindent {\bf Fig-19.} Probability distribution ($P(t_e)$) of 
first passage time ($t_e$) in three dimensions.
Different symbols correspond to different values of $\alpha$.
$\alpha=0.0 (\circ)$, $\alpha = 0.001(\bullet)$ and $\alpha=0.002(\ast)$. 
Here, $N_s=10^5$, $N_t=10^4$ and $r_e=25.0$.
\newpage

\setlength{\unitlength}{0.240900pt}
\ifx\plotpoint\undefined\newsavebox{\plotpoint}\fi
\sbox{\plotpoint}{\rule[-0.200pt]{0.400pt}{0.400pt}}%
%

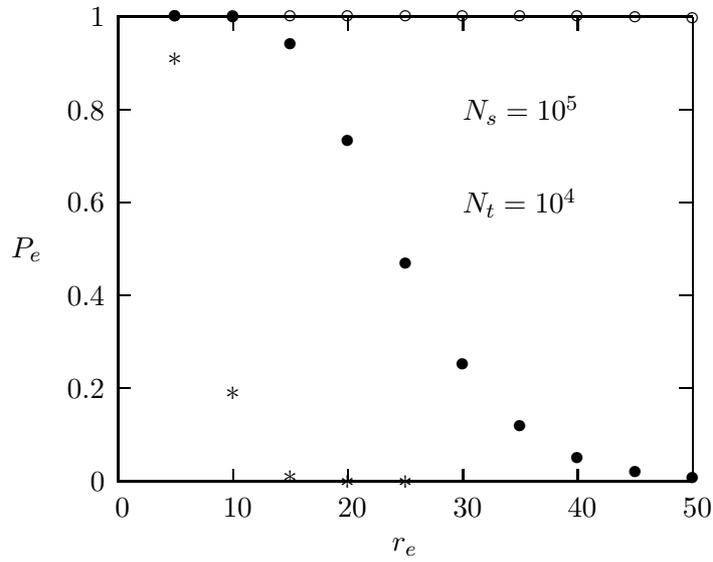
\noindent {\bf Fig-20.} Exit probability 
$P_e$ plotted against $r_e$ in three dimensions.
Different symbols correspond to different values of $\alpha$.
$\alpha=0.0 (\circ)$, $\alpha = 0.001(\bullet)$ and $\alpha=0.01(\ast)$. 
Here, $N_s=10^5$ and $N_t=10^4$.
\newpage
\end{document}